# Parallelism detection using graph labelling

P. N. Telegin, A. V. Baranov, B. M. Shabanov, and A. I. Tikhomirov

*Joint Supercomputer Center of the Russian Academy of Sciences - Branch of Federal State Institution Scientic Research Institute for System Analysis of the Russian Academy of Sciences, Leninsky prospect, 32a, Moscow, 119334, Russia*

**Abstract**

Usage of multiprocessor and multicore computers implies parallel programming. Tools for preparing parallel programs include parallel languages and libraries as well as parallelizing compilers and convertors that can perform automatic parallelization. The basic approach for parallelism detection is analysis of data dependencies and properties of program components, including data use and predicates. In this article a suite of used data and predicates sets for program components is proposed and an algorithm for computing these sets is suggested. The algorithm is based on wave propagation on graphs with cycles and labelling. This method allows analyzing complex program components, improving data localization and thus providing enhanced data parallelism detection.

**Introduction**

Increasing performance of computers requires the use multiprocessor and multicore architectures. This requires program to be parallel. When writing new programs typically parallel languages or parallel language extensions are used, as well as parallel libraries. Another way Iis to use automated tools for parallelization. This approach can really help when you are porting the code (e.g. to GPUs) or using legacy code. Parallelization can be automatic or semi-automatic. Semi-automatic parallelization is user-guided, i.e. a tool gives suggestions, and programmer decides what and how to parallelize. Among parallelizing convertors and compilers we can note the Cetus source-to-source parallelizer [1], system for automated parallelization SAPFOR [2], compiler for heterogeneous systems POLY-ACC [3], parallelizing compiler OSCAR [4] automatic parallelizer BERT-77 [5], automatic convertor of C code to multithreaded code YUCCA [6], automatic parallelization in MSVC [7].

These tools use the basic approach for parallelism detection in the application programs which is based on collecting information on program components [8]. Program properties encapsulated in program components can be used to determine dependencies and, hence, possibility of parallelization. This way was the Ratio automatic parallelizer is built [9]. Properties of program components always include sets of used data. Besides that, sets of predicates are formed to improve precision of parallelism detection. There are several approaches for parallelization, but the cornerstone is data dependencies [10]. Data dependencies set the limits for program transformations. To make more data transformations additional program properties are used like it is done in [11] or [12], where affine space transformations are utilized. The choice of these properties and their determination can be very specific for each tool. This article describes an approach for data use sets and algorithm for their determination that lease parallelism inhibitors.

**The model**

The program is represented as a directed graph $G=\{V, E\}$, Vertices ($V$) are program statements. edges ($E$) are transitions.

Each vertex has multiple labels describing properties of program statement. Here we consider properties important for parallelism detection.

Predicates are used both for improving parallelism both in static and dynamic analysis, as well for organizing speculative computations [13].

The sets bound variables (predicates) are assigned to nodes of program graph. We are considering three forms of predicates

1. $Pr1$: x = (n1:n2:n3)
2. $Pr2$: v1 <= y <=v2
3. $Pr3$: z = φ (S),

where x, y and z are variables
n1, n2, n3, v1, v2 are values (constants of variables expressions).

Pr1 defines exact values in the form (lower bound: upper bound: step)

Pr2 defines bound by which the variable x is limited.

Pr3 states that variable z is the result of operation in a given set. Here φ is an associative operation, like Π, Σ, *min*, *max*, etc. on the set S of variables.

Becides predicates there are the following labels assigned to the graph. To see that let's look on parallelism conditions of two operations ρ1 and ρ2. Note that operations can be complex like if-then-else, linear part of code, loop and iteration of loop or function. For each operation there are four sets of variables: $I(\rho), O(\rho), D(\rho), F(\rho)$.

$I(\rho)$ – all variables which values are used for execution of operation $\rho$.

$O(\rho)$ – all variables which values are changed as the result of operation $\rho$ executions.

$D(\rho)$ – all variables which values are changed at *every* executions of operation $\rho$. This needs to be explained. In the case of conditional operations like if-then-else sets of variables changed during execution of operation can be different depending of branch. The set $D(\rho)$ contains variable which are changed at every execution of $\rho$. Note that $D(\rho) \subseteq O(\rho)$.

Condition of parallelism for operations *p1* and *p2* (*p1* precedes *p2*) is the following [8]

$$I(p1) \cap O(p2) \cup O(p1) \cap (I(p2) \cup O(p2) \setminus D(p2)) \cup O(p1) \cap D(p2) \neq \emptyset \qquad (1)$$

Here [14]

$O(p1) \cap (I(p2) \cup O(p2) \setminus D(p2))$ is true (flow, forward) dependence

$I(p1) \cap O(p2)$ is backward (anti) dependence

$O(p1) \cap D(p2)$ is output (record) dependence

It is important to note that only true dependencies fundamentally limit program parallelization as was shown in classic work [15]. Other dependencies limitations can be avoided using program transformations. Our goal is to lease limitations set by true dependence and enhance parallelism that way.

While $I(\rho)$ and $O(\rho)$ sets are widely used and need no additional description, the $D(\rho)$ set may need some additional explanation. In Fig. 1 for simplicity, we mark each branch as a single operation. Each iteration of DO loop contains usage of variable TMP. As for conditional branch p3 the value of TMP is both in I(p3) and D(p3) sets. When we take a look at the loop we can see that TMP variable is changed not at every loop iteration. This way it impedes loop parallelization as it is not know from which iteration the value is taken, so loop-cariied dependencies may occur. Sometimes variables like this are added to the $I(p)$ set. We do not this because besides inhibition of concurrency this may result unnecessary costly data transfers.

First of all let's use another set of data $F(p)$

$F(\rho)$ – all variables which values are used to produce result after execution of operation $\rho$. These are variables which are not localized in operation $\rho$.

Usage of $F(\rho)$ set increases possibilities of data localization. Indeed, let's consider the following loop

After execution of the loop, it is generally not known from which iteration the TMP value should be taken. Of course, some program changes can save the needed iteration number, but this affects greatly on efficiency of parallel program especially on distributed memory systems. Presence of absence of variable TMP in the set $F(\rho)$ allows to make efficient decision.

**The Algorithm**

To calculate the *IODF* sets [8] and predicates the following graph labeling algorithm is used.

0. Vertices are labelled with IODF, according to usage of data by operations.
1. At the beginning the first statement s0 = V(0) and the last statement sn = V(n) are marked with token w.
2. Initialize data for all s ∈ V (I(s),O(s),D(s), F(s), Pr1(s), Pr2(s), Pr3(s))
3. Find next marked vertex si = V(i).
4. For vertex Vi for each forward transition from E find corresponding vertices VF ⊂ V.
5. For each corresponding vertex sj ∈ VF modify sets I(sj), O(sj) and D(sj) and sets of predicates. If any set was changed then mark vertex V(j) with token.
6. For vertex Vi for each backward transition from E find corresponding vertices VB ⊂ V.`
7. For each corresponding vertex sj ∈ VB modify the set F(sj). If any the set F(sj) was changed then mark vertex V(j) with token.
8. If any vertex is marked with a token then goto step 3. Otherwise, finish the algorithm.

The idea of the algorithm is wave propagation and it is similar to the Lee wave algorithm [16], Zhu/Ghahramani, Y.Wang et al label propagation [17], [18]. The difference is that it is generalized for graphs with cycles. As the wave propagates, it changes labels on nodes.

Example. Let's consider program in Fig. 1.

```
p1: for(I = 1; I<=N; ++I)
p2:    if(Condition(I))
p3:    {
           TMP = S(I)/2;
           A(I) = TMP;
       }
p4:    else{
           A(I) = S(I);
p5:    }
p6: }
```
Fig 1. Program with loop and if-then-else operations

In Fig.2 we consider propagation of data forming the *D* sets of variables. The corresponding graph is represented in Figures 2a-2h. On the right side of vertices you can see data from the *D* sets. Vertices *p1-p7* are marked only in Fig.2a.

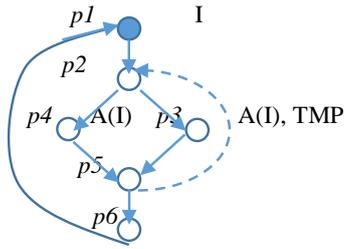

Fig. 2a

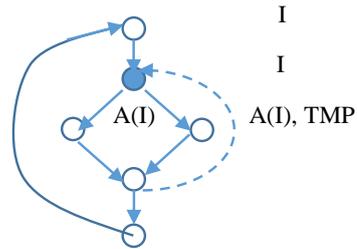

Fig. 2b

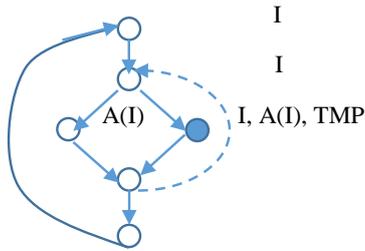

Fig. 2c

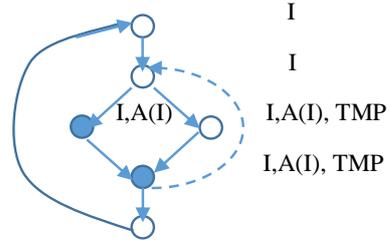

Fig. 2d

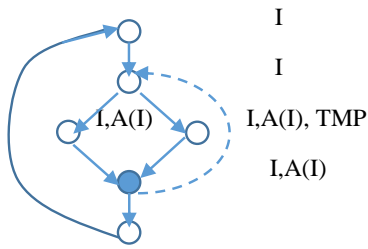

Fig. 2e

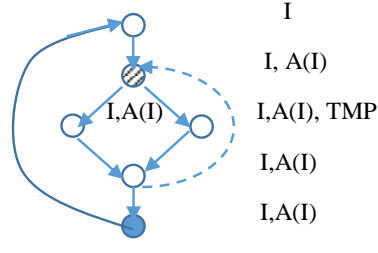

Fig. 2f

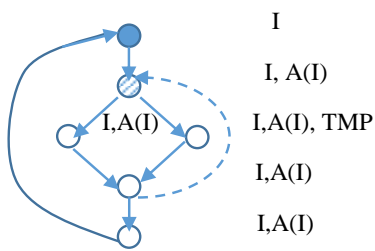

Fig. 2g

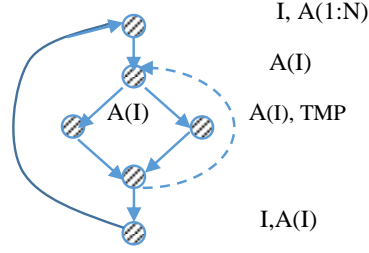

Fig. 2h

Fig. 2. Wave propagation

Initial labelling can be seen in Fig. 2a. Vertex *p1* is labelled with variable I (loop parameter), p3 is labelled with A(I) and TMP (complex operation consisting of two assignment statements), p4 is marked with A(I).

In Fig 2b we can see that the wave moved to p2, and label I has been replicated.

In Fig. 2 c the wave moved to branch p3, and it now has three labels.

In Fig. 2d we see that wave reached endif instruction, and it is marked with 3 labels as well.

Then the wave from p3 reaches endif (Fig 2e), and it is found that the wave comes without TMP variable label. So, TMP is removed from the *D* set of vertex p6.

After the if-then-else construction is processed, its *D* set is marked with I and A(I) variables (Fig. 2g). This is label not for condition but for the whole control structure.

In Fig 1h one can see the final result. As the wave make no changes on from p1 to p2, it stops. Extra data not belonging to statements or structures is removed. Note that vertex *p7* (ENDDO) contains data for one loop iterations, and label *p1* if marked for the whole DO loop.

**Example**

Let's see how the graph can be labelled, and how the resulting sets can help to enhance program parallelism.

```
v1: IF(K<0) THEN
v2:    K=-K
v3: ENDIF              K≥0
v4: DO I = 1, N
v5:    TMP = X(I) * Y(I)
v6:    Y(I) = Y(I+K) + TMP - 1.
v7: ENDDO                                              I=(1:N:1)
v8: S=0
v9: DO I = 1, N
v10:   IF(Y(I) > 0) THEN                               I=(1:N:1)
v11:      S = S + Y(I)
v12:   ELSE
v13:      S = S - Y(I)
v14:   ENDIF                            S = $\sum_{i=0}^{n} Y(I)$
v15: ENDDO                                             I=N+1
v16: PRINT *, S
```

Fig.3 Sample program

In Fig.4 you can find initial graph labelling. Only data related to operations is presented. Note, that you can see predicates at the vertices connected to those, which operations set the predicates values.

| Vertex | IODF | Pr |
|---|---|---|
| v1 | I={K} | |
| v2 | I={K}, O={K}, D={K} | K≥0 |
| v3 | | K≥0 |
| v4 | I={N}, O={I}, D={I} | |
| v5 | I={X(i),Y(I)}, O={TMP}, D={TMP} | I=(1:N:1) |
| v6 | I={K,Y(I+K),TMP}, O={Y(I)},D={Y(I)} | I=(1:N:1) |
| v7 | I={K,Y(I+K),TMP}, O={Y(I)},D={Y(I)} | I=(1:N:1) |
| v8 | O={S}, D={S} | |
| v9 | I={N}, O={I}, D={I} | S = 0. |
| v10 | I={Y(I)} | I=(1:N:1) |
| v11 | I={S,Y(I)},O={S}, D={S} | Y(I)>0 |
| | | S = $\sum$ Y(I) |
| v12 | | Y(I) ≤ 0 |
| v13 | I={S,Y(I)},O={S}, D={S} | I=(1:N:1) |

| | | $S = \sum -Y(I)$ |
|---|---|---|
| v14 | | |
| v15 | | |
| v16 | I={S,STODOUT}, O={STDOUT}, F={S} | |

Fig.4 Initial graph labelling

In Fig. 5 one can see the algorithm execution result. Note that date at v7 and v15 vertices (ENDDO) stands for iterations, and data for v4 and v9 stands for the whole loops.

| Vertex | IODF | Pr |
|---|---|---|
| v1 | I={K}, F={R, X(1:N), F=Y(1:N+ABS(K))} | |
| v2 | I={K}, O={K}, D={K}, F={R, X(1:N), F=Y(1:N+ABS(K))} | |
| v3 | I={K}, O={K}, D={K}, F={R, X(1:N), F=Y(1:N+ABS(K))} | $K \geq 0$ |
| v4 | I={N}, O={I}, D={I} F= F={N,K,X(1:N), Y(1:N+K)} | $K \geq 0$ |
| v5 | I={X(i),Y(I)}, O={TMP}, D={TMP}, F={N,K,X(1:N), Y(1:N+K)} | $I=(1:N:1)$ $K \geq 0$ |
| v6 | I={K,Y(I+K),TMP}, O={Y(I)},D={Y(I)} , F={N,K,X(1:N), Y(1:N+K)} | $I=(1:N:1)$ $K \geq 0$ |
| v7 | I={K,Y(I+K),TMP}, O={Y(I)},D={Y(I)} , F={N,K,X(1:N), Y(1:N+K)} | $I=(1:N:1)$ $K \geq 0$ |
| v8 | O={S}, D={S}, F={S} F={N,Y(1:N) } | |
| v9 | I={N,S,Y(1:N)}, O={I, Y(1:N), S}, D={I, Y(1:N), S}, F={N,S,Y(1:N)} | $S = 0.$ |
| v10 | I={Y(I)} , F={Y(I)} | $I=(1:N:1)$ |
| v11 | I={S,Y(I)},O={S}, D={S}, F={Y(I)} | $Y(I)>0$ $S = \sum Y(I)$ |
| v12 | F={Y(I)} | $I=(1:N:1)$ $Y(I) \leq 0$ |
| v13 | I={S,Y(I)},O={S}, D={S}, F={Y(I)} | $I=(1:N:1)$ $S = \sum -Y(I)$ |
| v14 | | $I=(1:N:1)$ $S = \sum +|Y(I)|$ |
| v15 | I={S,Y(I)},O={S,Y(I)}, D={S,Y(I)}, F={S} | $I=(1:N:1)$ $S = \sum_{i=0}^{n} |Y(I)|$ |
| v16 | I={S,STODOUT}, O={STDOUT}, F={S} | |

Fig.5. Result of the algorithm

Let's see what inhibitors in the sample program are loosened using our approach.

*F set*. Here we can see that in the first loop (corresponding vertices v4-v7) value of variable TMP is localized within iteration (cf. lastprivate clause in OpenMP). As it does not belong to the set F(v15), no action to save value from last iteration is needed.

*Predicate Pr2*. Value of variable K is greater or equal zero. So, there are no loop-carried dependences, and the loop becomes parallel.

*Predicate Pr3*. As for the second loop (v9-v15), it becomes equivalent to the statements

```
WHERE (Y(1:N) > 0) S = S + Y(I)
   ELSEWHERE   S = S - Y(I)
END WHERE
```

and this is reduction.

**Implementation**

The algorithm was implemented in the RATIO automatic parallelization system. 3 separate passes were implemented for computing:

- *I, O, D* sets
- *F* set
- *Pr1, Pr2, Pr3* sets

For computing *I,O,D* and *pr1,pr2,pr3* sets forward wave propagation is used. *F* set is computed via backward wave propagation.

In Fig.3 you can see function implementing forward wave propagation

```
void forward_p(void (*init)(void),void (*overture)(int i),
int (*action)(int j, int i), void (*underture)(int i),
void (*deinit)(void))
{
   int i,j,iter;
   my_flow[0]=1; // mark first statement with token
   n_flow=1;     // condition not finished
   init();       //
   for(iter=0;n_flow;++iter){
      n_flow=0;
      for(i=0;i<nins;++i){
         if(my_flow[i]){   // Wave already reached vertex i
            ++n_flow;
            my_flow[i]=0;
            overture(i);
            for(j=0;j<nins;++j){
               if(get_bit(flow_g[i],j)){   // process transitions
                  my_flow[j] = action(j,i) || my_flow[j];
               }
            underture(i);
         }
      }
   }
   deinit();
}
```
Fig. 3. C program implementing wave propagation algorithm.

The theoretical complexity of the algorithm is the number of operations multiplied be the number of loops. This is why wave algorithms are rarely used for graphs with cycles. Though this algorithm may require many passes, it is well suited for computer programs. Indeed, in the case of program graphs its complexity can be reduced to the number of operations multiplied be the maximum nested loop depth, this radically changes performance. This is due to the fact that for the most of computer languages program is executed in linear order with some control transitions, which are usually well-structured. More than that, while using this algorithm in the Ratio tool authors found no programs that required more than two passes regardless of loop nests depth.

**Conclusion**

In this article s suite of sets for parallelism detection is suggested. The suite includes four data usage sets and three predicates sets for each program component. Using combination of these sets enhances data localization, loosens parallelization inhibitors, and allows extra code transformations.

A wave graph labelling algorithm for graphs with cycles is suggested. This algorithm was implemented in the Ratio tool for automatic and user-guided parallelization. The algorithm shows acceptable performance on real program code.